\documentclass[twocolumn,amsmath,amssymb,10pt,superscriptaddress,a4paper,letterpaper,fleqn]{revtex4-1}
\usepackage{amssymb}
\usepackage{epsfig}
\usepackage{graphicx}
\usepackage{dcolumn}
\usepackage{array}
\usepackage{bm}
\usepackage{fancyheadings}
\usepackage{longtable}
\usepackage{multirow}
\usepackage{float}
\usepackage{afterpage}
\usepackage{color}

\setlongtables
\usepackage[breaklinks=true,linkbordercolor={1 1 1}]{hyperref}

\begin{document}
\setcounter{page}{1}

%%% **********************************************************************

\title{
%% Please do not remove the line below
\qquad \\ \qquad \\ \qquad \\  \qquad \\  \qquad \\ \qquad \\
%% Change title, authors, afiliation and type  your abstract
Combining Total Monte Carlo and Benchmarks for nuclear data uncertainty propagation on an LFR’s safety parameters}

\author{Erwin Alhassan}
\affiliation{Division of Applied Nuclear Physics, Department of Physics and Astronomy, Uppsala University, Uppsala, Sweden}

\author{Henrik Sj\"ostrand}
\email[Corresponding author: ]{henrik.sjostrand@physics.uu.se}
\affiliation{Division of Applied Nuclear Physics, Department of Physics and Astronomy, Uppsala University, Uppsala, Sweden}

\author{Junfeng Duan} 
\affiliation{Division of Applied Nuclear Physics, Department of Physics and Astronomy, Uppsala University, Uppsala, Sweden} 

\author{Cecilia Gustavsson} 
\affiliation{Division of Applied Nuclear Physics, Department of Physics and Astronomy, Uppsala University, Uppsala, Sweden} 

\author{Arjan Koning} 
\affiliation{Division of Applied Nuclear Physics, Department of Physics and Astronomy, Uppsala University, Uppsala, Sweden} 
\affiliation{Nuclear Research and Consultancy Group (NRG), Petten, The Netherlands}

\author{Stephan Pomp} 
\affiliation{Division of Applied Nuclear Physics, Department of Physics and Astronomy, Uppsala University, Uppsala, Sweden} 

\author{Dimitri Rochman} 
\affiliation{Nuclear Research and Consultancy Group (NRG), Petten, The Netherlands}

\author{Michael \"Osterlund} 
\affiliation{Division of Applied Nuclear Physics, Department of Physics and Astronomy, Uppsala University, Uppsala, Sweden}

\date{\today} 
%\received{8 March 2013; revised received XX June 2013; accepted XX September 2013}

\begin{abstract}
{Analyses are carried out to assess the impact of nuclear data uncertainties on $k_{eff}$ for the European Lead Cooled Training Reactor (ELECTRA) using the Total Monte Carlo method. A large number of Pu-239 random ENDF-formated libraries generated using the TALYS based system were processed into ACE format with NJOY99.336 code and used as input into the Serpent Monte Carlo neutron transport code to obtain distribution in $k_{eff}$. The $k_{eff}$ distribution obtained was compared with the latest major nuclear data libraries – JEFF-3.1.2, ENDF/B-VII.1 and JENDL-4.0. A method is proposed for the selection of benchmarks for specific applications using the Total Monte Carlo approach. Finally, an accept/reject criterion was investigated based on $\chi^{2}$ values obtained using the Pu-239 Jezebel criticality benchmark. It was observed that nuclear data uncertainties in $k_{eff}$ were reduced considerably from 748 to 443 pcm by applying a more rigid acceptance criteria for accepting random files.}
\end{abstract}
\maketitle

%%% DO NOT EDIT the following section enclosed by *****
%%% ***************************************************
%\lhead{ND 2013 Article $\dots$}
%\chead{NUCLEAR DATA SHEETS}
\rhead{Erwin Alhassan \textit{et al.}}
\lfoot{}
\rfoot{}
\renewcommand{\footrulewidth}{0.4pt}
%%% ***************************************************

%%% EDIT: the body of your text starts here, you may use as many \section, \subsection, \subsubsection
%%% \begin{figure}, \begin{tabular} and \begin{equations} as needed. Please note that each \begin{}
%%% must be closed with the corresponding \end{} and that section titles should be in capital
%%% letters. Current text should be eventually deleted.
\section{ INTRODUCTION}
{In recent times, there has been an increased use of nuclear reaction codes in the nuclear data evaluation process for the computation of cross sections, spectra and angular distributions required for a large variety of applications ~\cite{06IAEA}. The use of model codes in the evaluation of nuclear data offers several advantages, among them; the preservation of energy balance and coherence of partial cross sections together with total and reaction cross sections, the prediction of data for unstable nuclei and providing data where experimental data are unavailable. Where experimental data are available, they are used for fine tuning input parameters to model codes. The Total Monte Carlo (TMC) method, developed around TALYS by the Nuclear Research Group, Petten, has the capability of incorporating microscopic nuclear physics and macroscopic nuclear reactor design into one simulation scheme ~\cite{06Kon}. In this way, uncertainties can be assigned to macroscopic reactor parameters together with their sensitivity to uncertainties in specific cross-sections. The probability distribution for various reactor parameters can also be observed. TMC has been tested for a variety of systems ~\cite{07Dim}. In this paper, we applied the Total Monte Carlo approach in assessing the impact of Pu-239 cross section uncertainties on the full core 3-D SERPENT Monte Carlo model of the European Lead Cooled Training reactor(ELECTRA). ELECTRA is a 0.5 MW reactor proposed within the GENIUS (Generation IV research in universities of Sweden) framework ~\cite{06Wal}. It aims at developing technology necessary for safe and economic deployment of Generation IV nuclear reactors, in particular the lead cooled fast reactor. Since Pu-239 makes up of 30\% of the total mass of the fuel, its uncertainties are expected to impact significantly on the core behavior of the reactor. A novel approach for the selection of benchmarks for applications is present in this work. The selection of benchmarks for applications is usually done with 'by eye' judgment. This method is dependent on the experience of the user and therefore not suitable for the Total Monte Carlo methodology which lays emphasis on automation and reproducibility ~\cite{06Kon}. Also, an accept/reject criterion for reducing uncertainties in reactor macroscopic parameters based on limiting $\chi^{2}$ values is investigated using the Pu-239 Jezebel criticality benchmark.}

\section{METHODOLOGY}
\subsection{Simulations}
The input files used in this study are the SERPENT geometry input file developed at KTH, Sweden~\cite{06Wal} and 740 random ENDF files obtained from the TENDL project~\cite{06Kon}. The Pu-239 Jezebel benchmark MCNP input file obtained from the International Handbook of Evaluated Criticality Safety Benchmark Experiments ~\cite{06Bri} was converted into a SERPENT file for simulations. The random files were processed into ACE format using the Njoy99.336 processing code at 1200K.  Simulations were performed with the Serpent Monte Carlo code version 1.1.17 ~\cite{06Lepp} for the ELECTRA core at zero burnup with the absorber drums set at startup position. Criticality calculations were performed for 20 inactive and 500 $k_{eff}$ active cycles with 50,000 neutrons per cycle corresponding to 25 million histories resulting in a statistical uncertainty in $k_{eff}$ of 21.5 pcm. ENDF-6 formatted files for Pu-239 obtained from the latest major nuclear data libraries – JEFF-3.1.2, ENDF/B-VII.1 and JENDL-4.0 were processed in a similar way and criticality calculations performed. The results obtained are compared with mean of the $k_{eff}$ distribution of the random nuclear data files. 

\subsection{Fast TMC}
The fast TMC methology presented in ~\cite{06Dim} is used in this work. However, instead of using it to obtain short computation times it is used to obtain a very accurate and precise estimate of the uncertainty in reactor parameters due to nuclear data ($\sigma_{_{ND}}$). In TMC, ($\sigma_{_{ND}}$) is obtained by running N different ND files. The result is a spread in the data ($\sigma_{obs}$) that is both due to statistics ($\sigma_{stat}$) and due to nuclear data ($\sigma_{_{ND}}$): 
\begin{equation}
\sigma _{_{{\text{obs}}}}^2 = \sigma _{_{ND}}^2 + \sigma _{_{stat}}^2
\label{Yi}
\end{equation}
Since both the parameter, e.g, $k_{eff}$, and its statistical uncertainty, $\Delta k_{eff}$, is determined by most transport codes, we obtain a distribution of both $k_{eff}$ and $\Delta k_{eff}$ when we run with our N random files. The spread in the distribution of $k_{eff}$ is then $\sigma_{obs}$ and the $\Delta\overline{k_{eff}}$ is equal to $\sigma_{stat}$. Eq.~\ref{Yi} holds assuming that there is no correlation between statistics and nuclear data; the correlation between $k_{eff}$ and the corresponding statistical error was investigated and no correlation was observed. The methodology produces an accurate estimate of $\sigma_{_{ND}}$ if $\Delta\overline{k_{eff}}$ is accurately determined by the transport code. To test this we executed the serpent input file 740 times with constant nuclear data, but with different random seeds. If $\sigma_{obs}$ = $\Delta\overline{k_{eff}}$ then the transport code determines $\Delta k_{eff}$ correctly; in our case with SERPENT and ELECTRA and this was found to be true.
\vspace{-5mm}
\subsection{Selecting benchmarks for applications}
\begin{figure}[!htb] %[h]
\includegraphics[width=0.95\columnwidth]{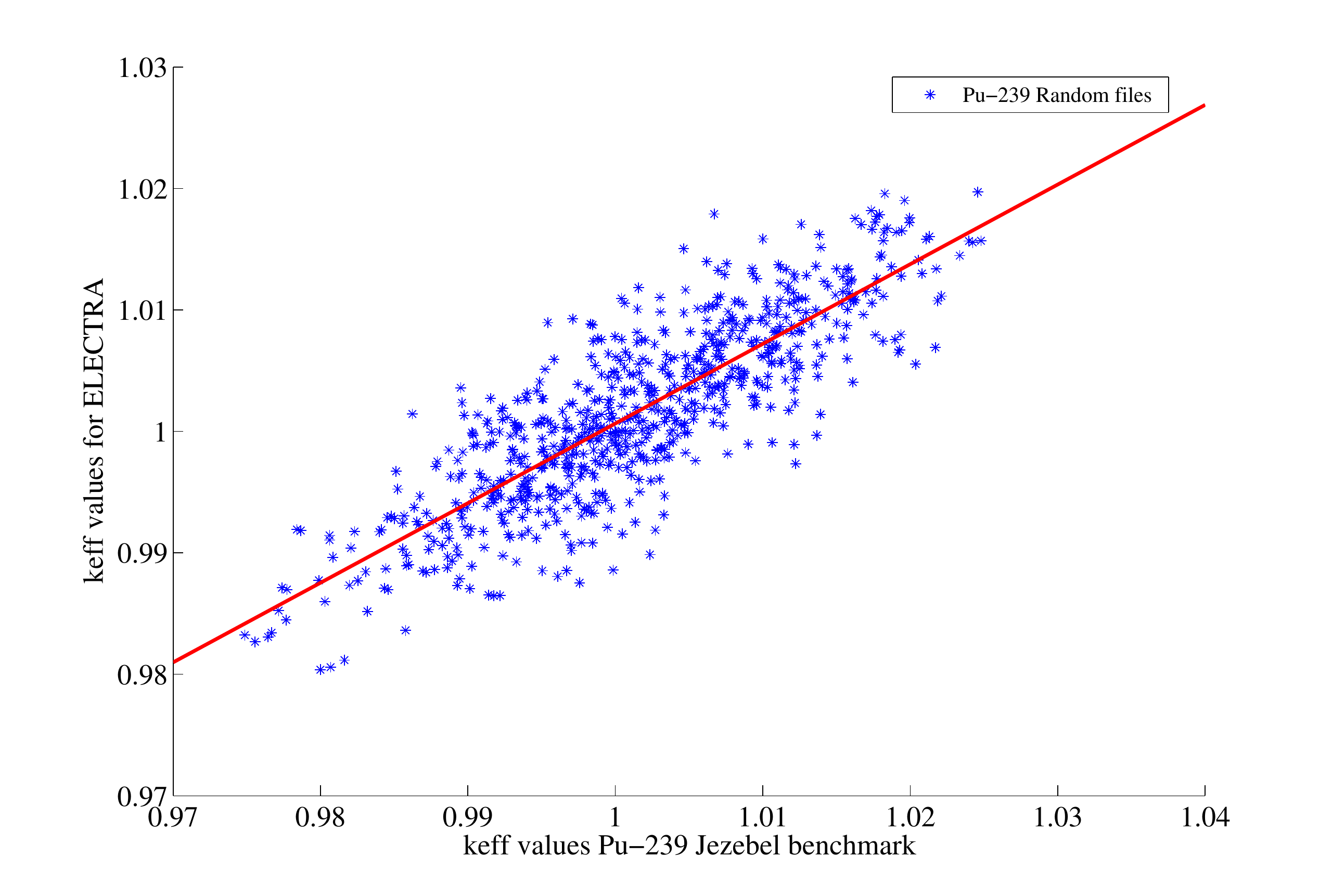}
\caption{Correlation between $k_{eff}$ determined for ELECTRA and  the Pu-239 Jezebel benchmark for Pu-239 random files.}
\label{fig1}
\end{figure}

Criticality benchmarks are used mostly for validation of calculational techniques in reactor calculations and for establishing minimum subcritical margins for operations with fissile material ~\cite{06Bri}. In the nuclear data evaluation process, benchmarking evaluated data against differential and integral data is very important. This is mostly done by selecting large sets of criticality, dosimetry, fusion, activation benchmarks among others. The selection varies from one application group to another as a results of different expertise, purpose of the evaluation and accessibility to benchmarks ~\cite{07Dim}. To solve this problem, we proposed the TMC methodology for the selection of benchmarks. First, random files obtained from the TENDL project are processed into usable format for reactor codes. Criticality simulations are performed for 1) the $i^{th}$ benchmark under consideration and for 2) the specific reactor system using same processed random files. A correlation test is then performed by plotting the $k_{eff}$ obtained for the reactor system against that of the benchmark. If a strong correlation exists between the two systems, this can be interpreted as a good representation of your system. This methodology has been tested only on the Pu-239 Jezebel benchmark and we plan to investigate this in more detail taking into consideration fusion and shielding benchmarks. The proposed method also opens up possibilities for assigning weights to each benchmark based on the goodness of fit or the correlation coefficient. In Fig. ~\ref{fig1}, a correlation between the $k_{eff}$ computed for ELECTRA and that of the Pu-239 Jezebel benchmark is presented. 
%\vspace{-5mm}
%
\subsection{Accept/reject criteria based on benchmarks}
An accept/reject criterion is investigated to study how limiting $\chi^{2}$ values for accepting random files could reduce nuclear data uncertainties in safety parameters. Simulations were performed for Pu-239 random files ~\cite{06Kon} and with Pu-239 Jezebel criticality benchmark and with the ELECTRA Serpent Monte Carlo code input file. Limiting chi squares were set and corresponding nuclear data uncertainty calculated.In Fig.~\ref{fig2}, we present a plot of limiting chi square verses Pu-239 random files index. 
%The goodness number, $\chi^2$ is calculate: 
%\begin{equation}
%\chi^{2}_{i}=\frac{(C_{i}-E)^{2}}{C_{i}}
%\label{chi01} 
%\end{equation}
%Where $C_{i}$ is the calculated $k_{eff}$ for the the $i^{th}$ random %file and $E$ is the nominal value for the benchmark, in this case %1.0000.
%
%\begin{figure}[h] %[h]
%\includegraphics[width=0.95\columnwidth]{Limiting_chi2_1.pdf}
%\caption{Limiting chi square against Pu-239 random file index.}
%\label{fig1}
%\end{figure}
%
\section{RESULTS}
\begin{figure}[h] %[h]
\includegraphics[width=0.95\columnwidth]{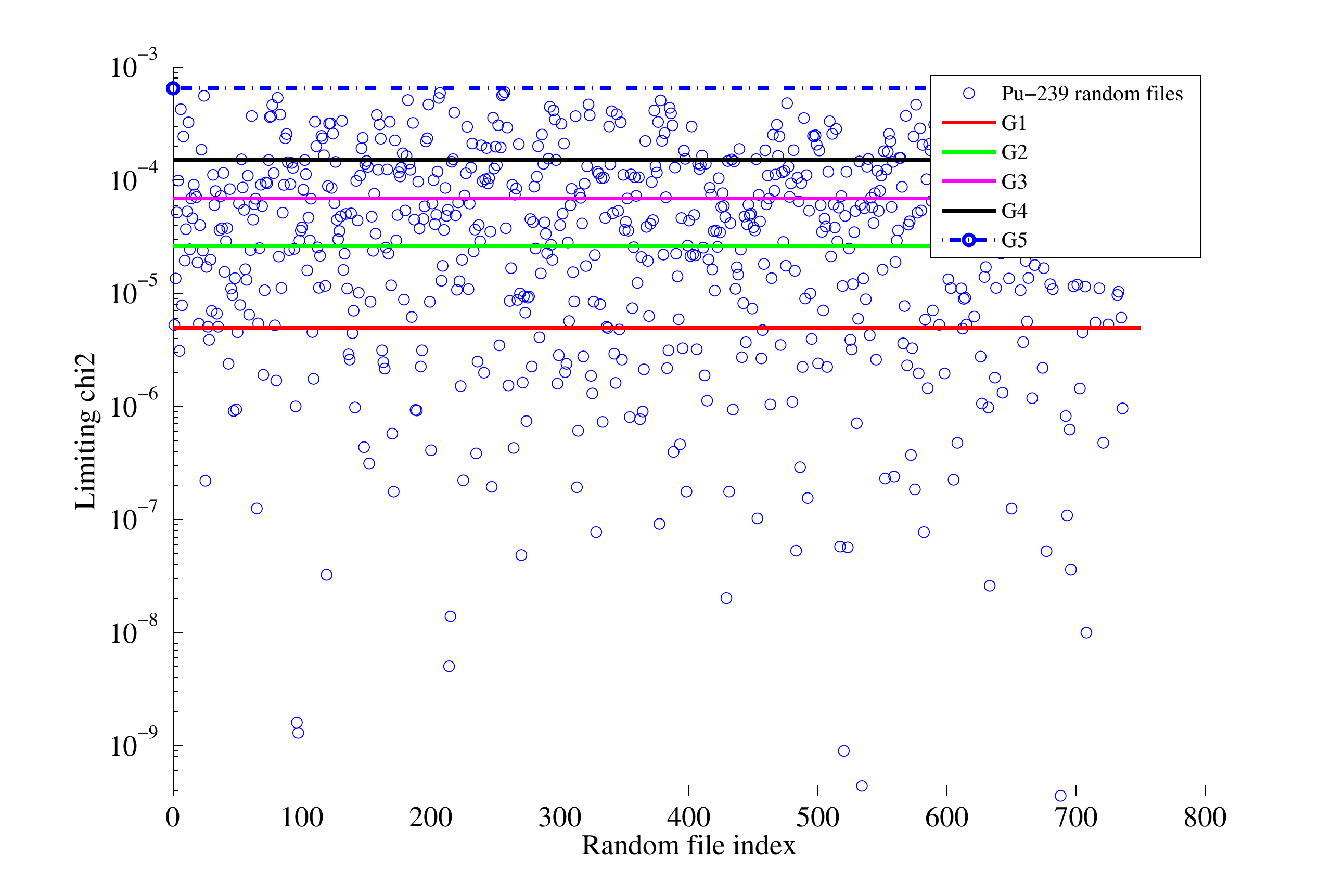}
\caption{Limiting chi square against Pu-239 random file index.}
\label{fig2}
\end{figure}

\begin{figure}[htb] %[h]
\includegraphics[width=0.95\columnwidth]{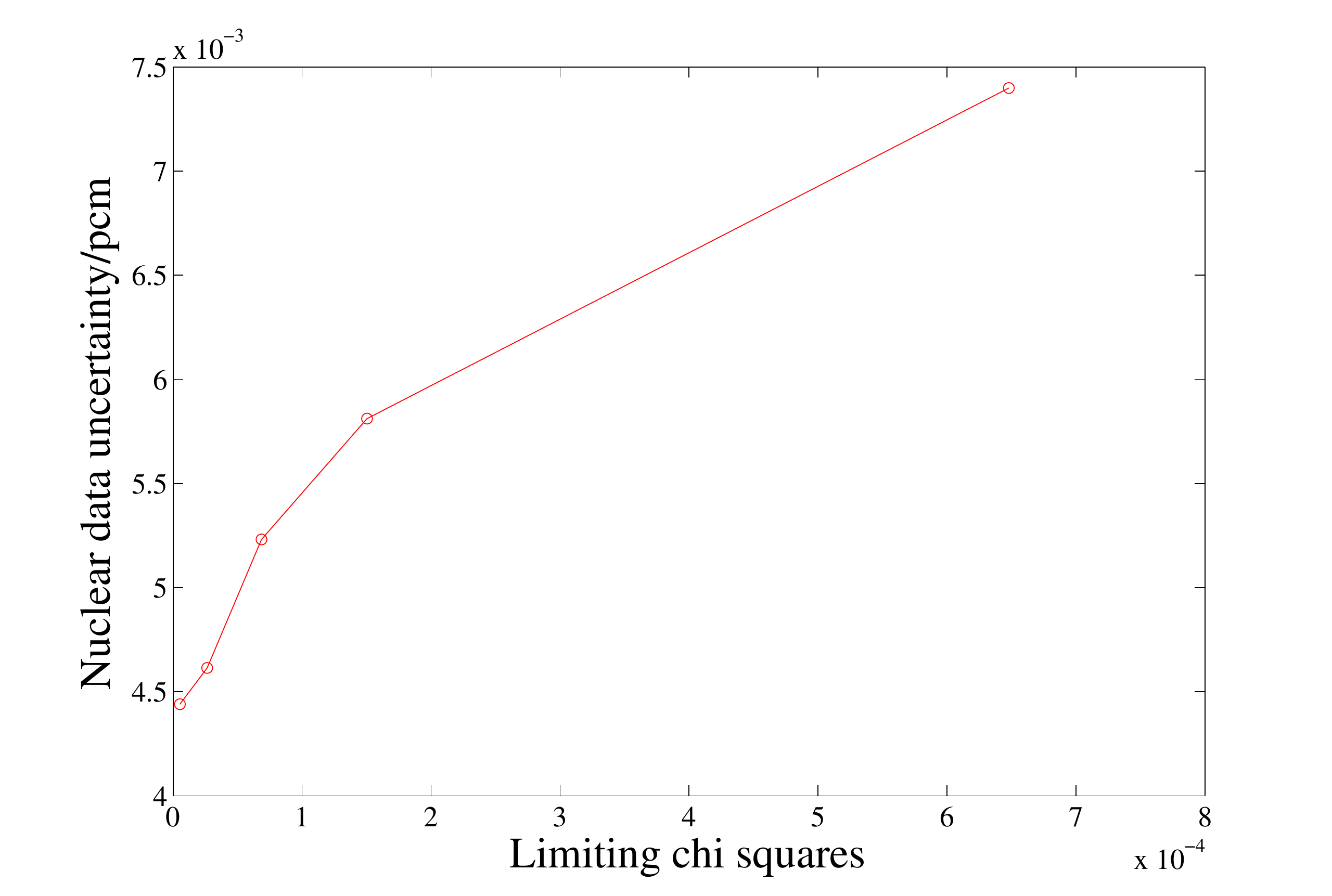}
\caption{Nuclear data uncertainty against limiting chi square using the Pu-239 Jezebel benchmark and Pu-239 random files.}
\label{fig3}
\end{figure}

A distribution in $k_{eff}$ was obtained for 740 Pu-239 random ENDF files and the mean compared with results from other nuclear data libraries. The JEFF-3.1 (the reference library) temperature dependent cross-section libraries were maintained for all other isotopes except for Pu-239 files which was changed with files from TENDL, JENDL-3.1.2, JENDL-4.0 and END/B-VII.1. It was observed that the mean value of of the random files $1.001540\pm0.00021$ compared favorably with the other nuclear data libraries: ENDF/B-VII.1 was $1.00067\pm0.00021$, JEFF-3.1.1 was  $1.00023\pm00022$, JENDL-4.0 was $1.00043\pm0021$. This can be attributed to the contribution of the well known fission cross section of Pu-239 which has a relatively high impact on the effective multiplication factor. The uncertainty due to nuclear data was determined and found to be $745\pm19$ pcm.
%
%Fig.2 shows a histogram of $k_{eff}$ distribution for Pu-239 random files compared with $$k_{eff}$ from other nuclear data libraries.
%
%\begin{figure}[] %[h]
%\includegraphics[width=0.95\columnwidth]{keff_Pu-239_2.pdf}
%\caption{$k_{eff}$ distribution for Pu-239 random files.}
%\label{fig2}
%\end{figure}
%\includegraphics[width=0.95\columnwidth]{k-eff distribution for Pu-239.eps}
%
In Fig.~\ref{fig2}, we present a plot of limiting $\chi^{2}$ values against the random file index. Nuclear data uncertainties were determined for each limit and the results is illustrated in Fig.~\ref{fig3}.

The correlation between $k_{eff}$ computed for ELECTRA and for the Pu-239 Jezebel benchmark is shown in Fig.~\ref{fig1}. It can be seen from the plot that there exists a strong correlation between the two systems implying that, the Pu-239 Jezebel benchmark is a good representation of ELECTRA. The Pearson correlation coefficient which is a measure of the strength of the relationship  between the two systems was computed and found to be R=0.84, further underscoring the strong correlation observed. For the computation of $\chi^{2}$ for a large set of benchmarks, we propose that a weighted factor ($w_i$), which should be equal to the absolute value of the correlation coefficient ($R_i$) be set for the $i^{th}$ benchmark. We plan to test this methodology on a large set of benchmarks and for different nuclides. The approach has however been applied Am-241 in burnup evaluation of ELECTRA ~\cite{0Hen}. It was observed that, a 25\% reduction in nuclear data uncertainty was achieved after a setting a more rigid acceptance criteria. 
In Fig.~\ref{fig3}, we present the nuclear data uncertainty against limiting $\chi^{2}$. it can be observed from the figure that by setting a minimum limiting $\chi^{2}$ of  5.31e-06, we were able to reduce the uncertainty in nuclear data from 748 to 443 pcm. This therefore implies that, by implementing an accept/reject criterion in the Total Monte Carlo method for accepting random files, nuclear data uncertainty can be reduced significantly. 

%\includegraphics[width=0.95\columnwidth]{ND uncertainty with limiting chi squares.pdf}
%\vspace{-5mm}
\section{ CONCLUSIONS}
A study has been carried out by using benchmarks within the Total Monte Carlo methodology on nuclear data uncertainty propagation for the European Lead Cooled Training Reactor (ELECTRA). It was observed that, nuclear data uncertainties could be reduced considerably in criticality calculations after introducing an accept/reject criterion based on integral benchmarks into the TMC calculation  chain.  A method for the selection of benchmarks with TMC was also proposed which opens up possibilities for assigning weights to benchmarks. 

\end{document}